\documentclass{aa}  
\usepackage{graphicx}
\usepackage{natbib}
\bibpunct{(}{)}{;}{a}{}{,}
\newcommand{\Msun}{\mbox{$M_{\odot}$}}

\newcommand{\vinf}{\mbox{$v_{\infty}$}}

\newcommand{\mdot}{\mbox{$\dot{M}$}}

\begin{document}
   \title{Implications of the metallicity dependence of Wolf-Rayet winds}
   
   \author{John J. Eldridge
          \inst{1}
          \and
          Jorick S. Vink\inst{2}
          }

   \institute{Department of Physics and Astronomy, Queen's University Belfast,
Belfast, BT7 1NN, Northern Ireland, UK\\
              \email{j.eldridge@qub.ac.uk}
         \and
             Astrophysics Group, Lennard-Jones Laboratories, Keele University, Staffordshire, ST55BG, UK\\
             \email{jsv@astro.keele.ac.uk}
             }

   \date{}

  \abstract{}
{Recent theoretical predictions for the winds of Wolf-Rayet stars indicate that their mass-loss rates scale with the initial stellar metallicity in the local Universe. 
We aim to investigate how this predicted dependence affects the models of Wolf-Rayet stars and their progeny in different chemical environments.}
{We compute models of stellar structure and evolution for Wolf-Rayet stars for different initial metallicities, and investigate how the scaling of the Wolf-Rayet 
mass-loss rates affects the final masses, the lifetimes of the WN and WC subtypes, and how the ratio of the two populations vary with metallicity.}
{We find significant effects of metallicity dependent mass-loss rates for Wolf-Rayet stars. For models that include the scaling of the mass-loss rate with initial metallicity, all WR stars become neutron stars rather than black holes at twice the solar metallicity; at lower $Z$, black holes have larger masses. 
We also show that our models that include the mass-loss metallicity scaling closely reproduce the observed decrease of the relative 
population of WC over WN stars at low metallicities. }
{}

\keywords{stars: evolution -- stars: Wolf-Rayet  -- stars: mass loss -- stars: early-type -- stars: winds, outflows}

\maketitle

\section{Introduction}
Massive stars are dominant sources of energy and nucleosynthesis products for the interstellar medium (ISM). This input is provided during their entire lives via stellar winds, followed by spectacular deaths as supernovae and long-duration gamma-ray bursts (GRBs). Because of their intrinsic brightness, they can be identified individually in extragalactic galaxies, whilst their integrated emission dominates the spectra of distant galaxies in the high redshift Universe. Wolf-Rayet (WR) stars are bare helium stars, with spectra that are characterised by broad emission lines, and a lack of hydrogen (H), which is a result of mass loss. Stellar winds continue to enrich the ISM during the WR phase, where the objects turn into important sources of helium (He), nitrogen (N), carbon (C), oxygen (O), and other elements in our Universe. In addition, the N-rich WN and C-rich WC stars are the favoured progenitors of type Ibc supernovae due to their lack of H \citep{EW88}; more recently they have become suspect to be the progenitors of long-duration GRBs \citep{grbprog}.

Nonetheless, WR stars remain somewhat of an enigma. Measuring the exact mass-loss rates (\mdot) and other stellar parameters is particularly challenging, as WR atmospheres are not in local thermodynamic nor hydrostatic equilibrium, which prevents a direct comparison of surface temperatures and radii with stellar models. WR winds are optically thick and inhomogeneous, and wind-clumping has led to severe overestimations of their mass-loss rates by factors of about three \citep{HK98,NL00}. The intricacies of clumping and their effect on measurements of WR mass-loss rates remain a source of uncertainty today.

Despite these uncertainties, we have significantly advanced our understanding of WR stars, as it has become increasingly clear that WR winds are driven by radiation pressure \citep{LA93,Hill03,GH05}, and that WR mass loss depends on their initial metallicity ($Z$). In particular, \citet{WRZscale} presented observational evidence for a WR $\mdot - Z$ relation for WC stars, while \citet{vink2005} recently showed how $\mdot$ is predicted to vary with initial metallicity for late-type WC as well as WN stars. 

In this article, we investigate the implications of this $\mdot - Z$ scaling for WN and WC stars. First we describe the construction of our stellar models and the mass-loss scheme we have implemented (Sect.~\ref{construction}). In Sect.~\ref{comp}, we discuss how our models compare to models of \citet{mm2005}. We subsequently predict final masses (Sect.~\ref{final}) and WR lifetimes (Sect.~\ref{wrlifetimessection}), and compare the model results to the observed WC/WN ratio at various metallicities (Sect.~\ref{ratio}). We finally check the sensitivity of our results to potential remaining uncertainties in the exponent of the scaling of WR mass-loss rates, before we conclude in Sect.~\ref{concl}.

\section{Construction the Stellar Models}
\label{construction}

\begin{figure}
\includegraphics[angle=0, width=84mm]{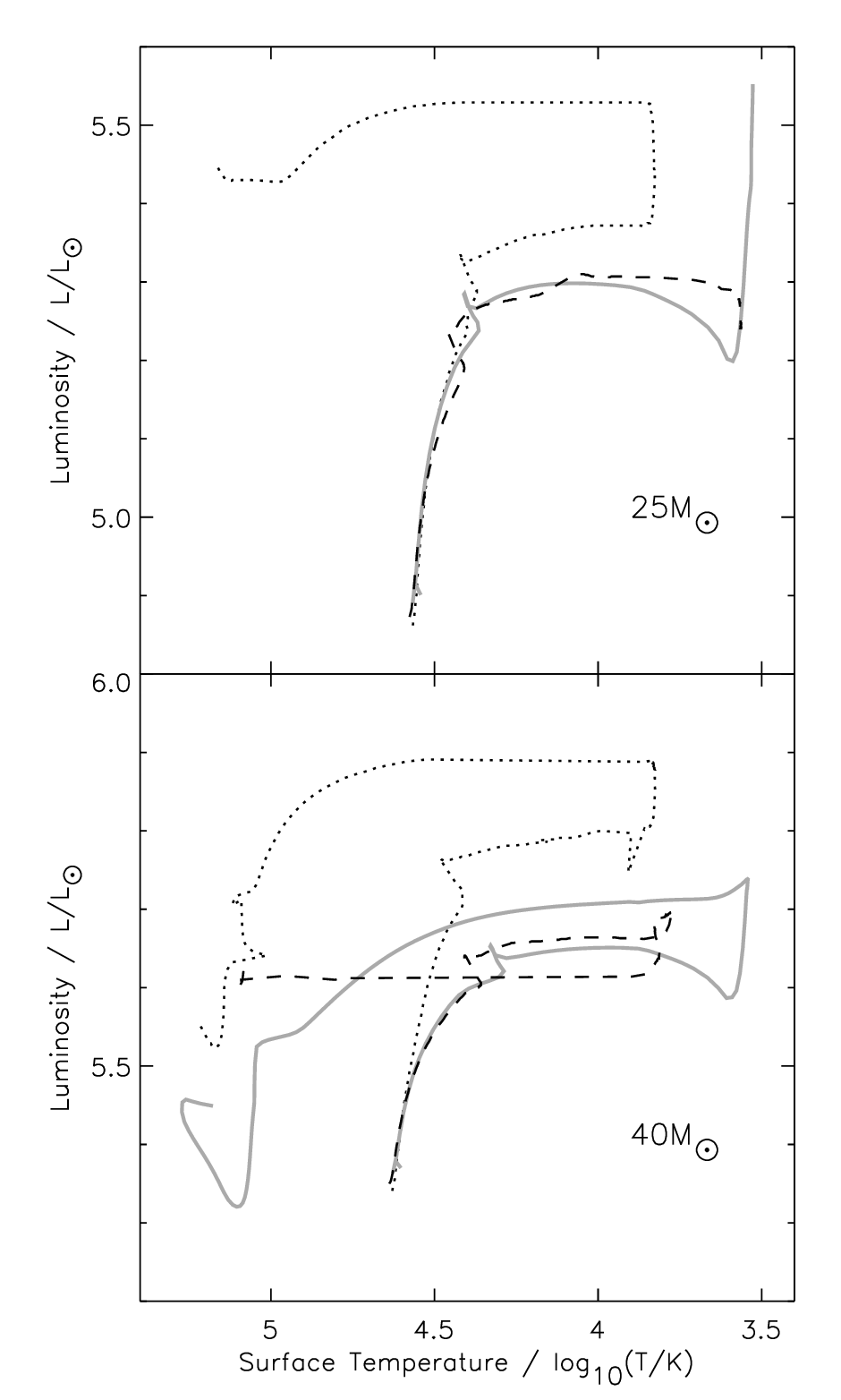}
\caption{Hertzsprung-Russell diagram comparison between our solar metallicity models and those from \citet{mm2003}. Our models are represented by the solid grey lines, the other lines are taken from \citet{mm2003}. The dashed lines indicate non-rotating models, while the dotted lines represent an initial rotational velocity of $300~{\rm km \, s^{-1}}$. The stars in the upper panel have an initial mass of $25\Msun$, while those in the lower panel have an initial mass of $40 \Msun$.}
\label{hr1}
\end{figure}

\begin{figure}
\includegraphics[angle=0, width=84mm]{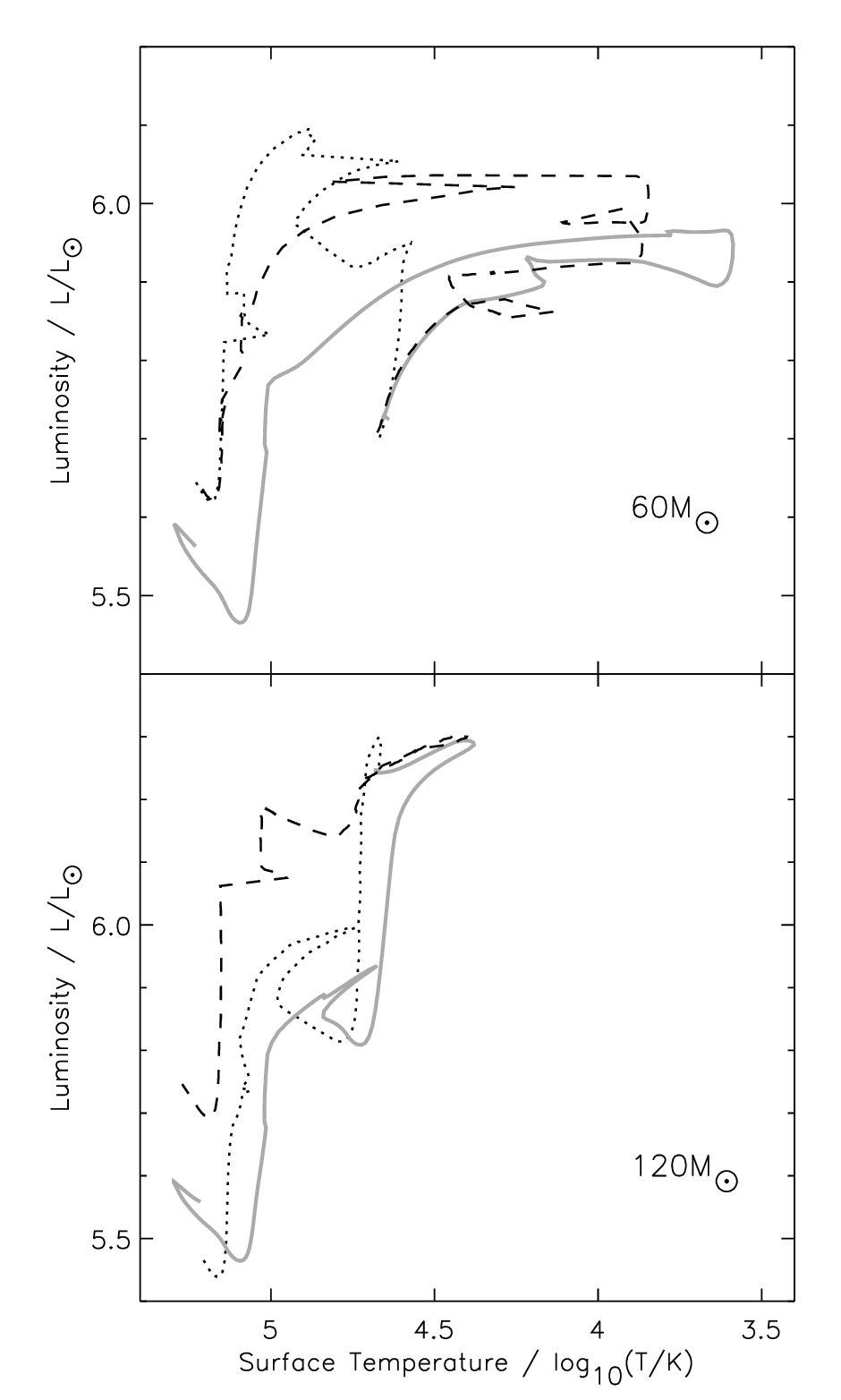}
\caption{Same as Fig.~\ref{hr1} for initial masses of $60 \Msun$ (upper panel) and $120\Msun$ (lower panel).}
\label{hr2}
\end{figure}

Our stellar models are produced with the Cambridge stellar evolution code, STARS, originally developed by \citet{E71} and updated most recently by \citet{P95} and \citet{E03}. Further details can be found at the code's home page \footnote{\texttt{http://www.ast.cam.ac.uk/$\sim$stars}}. The models are similar to the overshooting models described in \citet{ETsne}. The differences are as follows: the initial masses are 25, 30, 35, 40, 45, 50, 55, 60, 70, 80, 100 and $120{\rm M}_{\odot}$. $Z$, the initial metallicity, takes the values 0.001, 0.002, 0.003, 0.004, 0.008, 0.02 and 0.04, i.e. the range is $\frac{1}{20}Z_{\odot}$ to $2 Z_{\odot}$. The metallicity comprises scaled-solar abundances. In constructing the stellar models, we assume ``solar metallicity'' to correspond to a metal mass fraction $Z=0.02$ and scale the mass-loss rates from this point. The most recent analysis of the Sun's composition \citep{someone} suggests that this metallicity may need to be revised to a value as low as $Z=0.012$. Of course the metallicity of the solar neighbourhood may not exactly equal $Z_{\odot}$ itself, and it is the metallicity of the solar neighbourhood for which mass-loss rates have been measured and calibrated against. Furthermore, the relative scale is more important than the absolute value, and the uncertainties related to the value of the initial metallicity are well within the uncertainty of WR mass-loss rates.

The mass-loss scheme is the most relevant factor in constructing the stellar models. For OB stars, we use the mass-loss predictions of \citet{VKL2001}\footnote{An IDL routine to compute the mass-loss rate as a function of stellar parameters is publicly available at \texttt{http://www.astro.keele.ac.uk/$\sim$jsv/}} which include a metallicity scaling. For all other pre-WR phases, we employ the rates of \citet{dJ} scaled with metallicity by a factor of $(Z/Z_{\odot})^{0.5}$. When the star becomes a WR star ($X_{\rm surface}<0.4$, $\log(T_{\rm eff}/K)>4.0$), we use the rates of \citet{NL00}. It is the scaling of these rates with $Z$ that produces different sets of models. The scaling we apply depends both on the WR star type and the initial metallicity of the star. The WR star type is determined as follows. The star is initially a WNL star, and once $X_{\rm surface}=0$, the star becomes a WNE star. To determine the switch from WNE to WC star, we define the following parameter, $\zeta = (x_{\rm C}+x_{\rm O})/y$, where $x_{\rm C}$, $x_{\rm O}$ and $y$ are the surface number fractions of C, O and He. Following \citet{mm1994}, the star becomes WC when $X_{\rm surface}=0$ and $\zeta > 0.03$. In a similar vein, the switch to a WO star occurs when $\zeta > 1.0$, but we only encounter the WO phase when we do not include the $\mdot-Z$ scaling. Since WO lifetimes are much shorter than those of the other WR phases, we do not distinguish WO stars from WC stars.

With the WR type defined, we apply the $\mdot$ scaling factor, $(Z/Z_{\odot})^{x}$, where $x$ is shown in Table~\ref{schemetable}. When $Z<0.002$ the scaling factor changes to $(Z/0.002)^{x_{2}} \times (0.002/Z_{\odot})^{x_{1}}$ to achieve the correct scaling. $x_{1}$ is the exponent when $Z>0.002$ and $x_{2}$ is the exponent when $Z<0.002$. In scheme A, \mdot\ does not scale with $Z$. The preferred values of our $\mdot-Z$ dependence (scheme B) are taken from the study of late-type WN and WC $\mdot$ predictions of \cite{vink2005}, where the exponents come from those predicted for the wind momentum ($\mdot$ $\vinf$) dependence on $Z$. In other words, the underlying assumption for scheme B is that the wind velocity of WR stars does not vary with $Z$, as we anticipate a potential $Z$-dependence of $\vinf$ to be relatively weak. The reason is that the line force due to iron lines is predominantly found to determine the mass-loss rate, whilst the CNO elements set \vinf\ \citep{VKL99,PSL00}. Because these CNO abundances are largely determined by ``self-enrichment'', the \vinf\ versus initial stellar metallicity dependence (which scales with Fe) is expected to be even weaker for WR stars than it is for 
OB stars. Nonetheless, to accommodate for a potential metallicity dependence of the WR wind velocity, we include scheme C, where the exponents of scheme B have been slightly reduced. This scheme will be used for comparison later as to estimate the importance of the potential wind velocity metallicity dependence on WR models. We finally include scheme D to relate our computed exponents to the commonly quoted constant exponent of 0.5.

\begin{table}
\caption{Scaling of the WR mass-loss rates with metallicity. The values represent the exponent of 
$\dot{M}(Z)=\dot{M}(Z_{\odot})\big(\frac{Z}{Z_{\odot}}\big)^{x}$.}
\label{schemetable}
\begin{tabular}{lccccc}
\hline
	  &$Z$		& Scheme A& B & C & D\\
\hline
WN	  &--		& 0.0	&0.85 & 0.69 & 0.5\\
WC	  &$>$0.02	& 0.0	&0.4  & 0.3  & 0.5\\
	  &$<$0.02	& 0.0	&0.66 & 0.56 & 0.5\\
	  &$<$0.002	& 0.0	&0.35 & 0.25 & 0.5\\
\hline
\end{tabular}
\end{table} 

\section{Model comparison}
\label{comp}

We compare our stellar models to other contemporary models of WR stars. 
We do this for four initial masses in Figs.~\ref{hr1} and \ref{hr2}, using the non-rotating 
and rotating models of \citet{mm2003} and \citet{mm2005}, the Geneva models. 
The surface temperatures have not been corrected for 
the non-equilibrium effects of the WR atmosphere as described in \citet{mm2005}.
Although stellar rotation is inherently a 3D process, the Geneva models include the effects of stellar rotation by using approximations that can be 
included within a 1D evolution code. We note that the Geneva tracks end after core helium burning, while ours progress onto core carbon burning. 
%This leads to a minor difference in that our tracks being slightly greater in extent.

Comparing the $25\Msun$ models in Fig.~\ref{hr1}, we find that the non-rotating Geneva model agrees well with our model, although the Geneva track does not include the RSG phase. The rotating model differs significantly in that the star ends its evolution as a WR star. 
This is due to a combination of rotation enhancing the mass-loss rates, rotation increasing the main-sequence lifetime leading to extra mass-loss, 
and a larger helium core due to the rotational mixing. The $40\Msun$ models all end as WR stars. However, the Geneva tracks never become cool red supergiants as ours, and 
the rotating model produces more luminous stars due to enhanced energy production as a result of rotational mixing. 
The final masses of both types of Geneva models are a few $\Msun$ more massive than ours. Similar results are found in Fig.~\ref{hr2} for the $60\Msun$ models, but 
in addition the rotating model never moves over towards the red side of the HR-diagram. 
For the $120\Msun$ models, stars all remain in the blue part of the HR-diagram for their entire evolution. 
The rotating star in this cases achieves the same masses as in our model.

Overall the differences are relatively minor. They can be explained by two important differences between our models and those of \citet{mm2003}. 
Most details of the evolution codes are however very similar. For example, both include a small amount of convective overshooting. 
While both models employ the same base mass-loss rates of \citet{VKL2001}, \citet{dJ} and \citet{NL00}; their use is different. In our models we use all of the rates unmodified as detailed above. The Geneva models using the \citet{VKL2001} rates, which are for non-rotating stars as a basis to reduce the empirical rates to account for the rotation of the observed stars. They then enhance the mass-loss rate according to the rotation rate of the model. This leads to the mass-loss rate for their non-rotating models 
being less than those of our models, whilst the mass-loss rates employed for their rotating models are very similar to those of our models, as we 
use unaltered mass-loss rates.
The above implies that our models are similar to the rotating Geneva models, except that we do not consider the effects of rotational mixing. 
The extra mixing of hydrogen into the stellar core extends the main-sequence lifetime for rotating stars, which leads to smaller 
post main-sequence stellar masses. In addition, mixing leads to greater helium core masses, so that larger WR star masses are obtained in 
the Geneva models than in ours.
The Geneva non-rotation models have a higher minimum initial mass for WR stars, $37\Msun$,  compared to our $27\Msun$, and shorter WR lifetimes. 
While the rotating models have a lower minimum initial mass of $22\Msun$ and longer lifetimes. It is difficult to make a comparison at other metallicities as there 
are no contemporary non-rotating Geneva models at lower metallicities. Nonetheless, similar trends are found comparing the rotating models to our 
scheme A models. While the scheme B models differ by producing much greater final masses than do the same Geneva models.

In summary, the models we present are non-rotating models that have not been modified in anyway to account for rotation as in \citet{mm2003}. Importantly we do not consider how rotation introduces enhanced mixing and we do not alter the mass-loss rates to account for the rotation of the stars.

\section{Final WR star masses}
\label{final}

\begin{figure}
\includegraphics[angle=0, width=84mm]{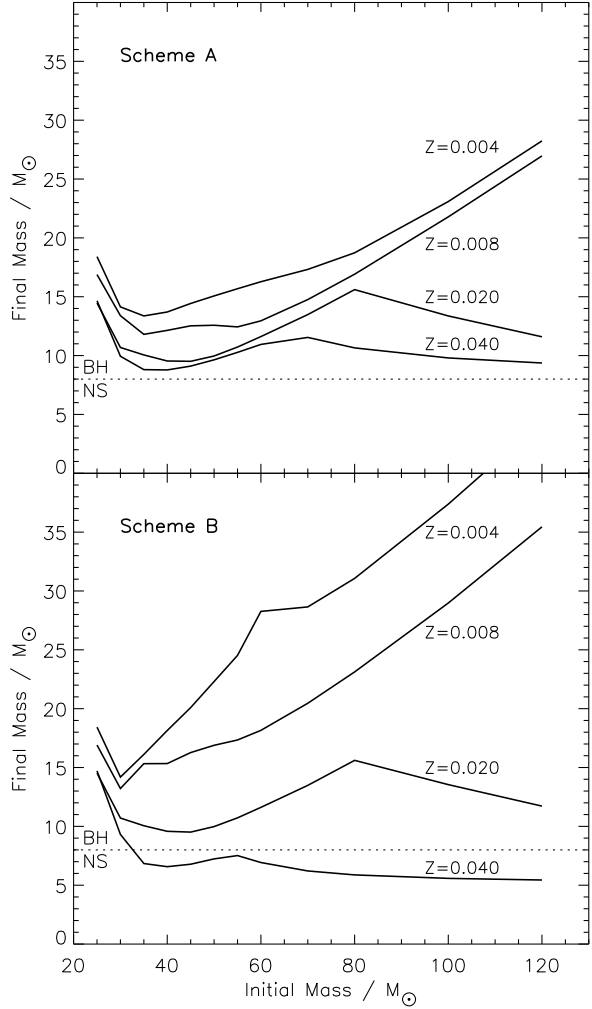}
\caption{The final masses from our models versus the initial masses at four different metallicities. 
The upper panel shows the predictions according to scheme A, where no WR scaling has been applied. The lower panel denotes predictions according to scheme B where the WR scaling is applied.}
\label{mass}
\end{figure}

We first consider how relevant aspects of the stellar models are affected by the scaling of the WR mass-loss predictions. The final mass of our stars is directly determined by the amount of mass lost over their lifetimes. We present our final mass predictions in Fig.~\ref{mass}, where we show results at four metallicities: solar, twice solar, two fifths solar and one fifth solar. In the upper panel for scheme A we see that lower metallicity WR stars are more massive due to the reduced pre-WR mass loss so the initial WR masses are larger. In the solar case, stars with initial masses $\ge$ $27M_{\odot}$ become WR stars. As the initial mass increases, so does the final mass up to a peak at $80M_{\odot}$, after which, the final mass again decreases. This is due to the time spent as a red supergiant becoming shorter, and effectively, the stars remain hot for their entire lifetimes, experiencing strong mass-loss during their main-sequence lives leading to lower masses.

Comparing these results to the lower panel of Fig.~\ref{mass} we find large differences. At twice solar metallicity the masses are decreased by a few solar masses. While at lower metallicities the final masses are increased by the same factor two fifths solar. Furthermore, the change in final mass from solar to these LMC-type models are similar to those predicted by \citet{WRZscale}. At the lowest metallicity here the effects are greatest and we find the masses to be larger by around 10$M_{\odot}$.
 
The effects on the final masses of WR stars are most relevant for the prediction of the masses of the compact remnants. If we assume that stars with He cores greater than $8M_{\odot}$  produce black holes at the end of their lives \citep{H03}, we find that by including the WR scaling at twice solar metallicity, all compact remnants are predicted to be neutron stars rather than black holes. On the other hand, at lower $Z$, the masses of black holes are predicted to be much larger with the scaling. In other words, the compact remnant mass function shows a strong $Z$-dependence.

\section{WR star lifetimes}
\label{wrlifetimessection}

The lifetimes of our WR models depend on the mass-loss rate in two ways. First, \mdot\ determines the {\it total} lifespan of the WR star by affecting the total stellar mass and thus the vigour of the nuclear reactions in the core. Second, stronger mass loss strips mass quicker off the stellar surface, leading to the exposure of the products of later nuclear reactions, resulting in the appearance of different WR subtypes. 

\begin{figure}
\includegraphics[angle=0, width=84mm]{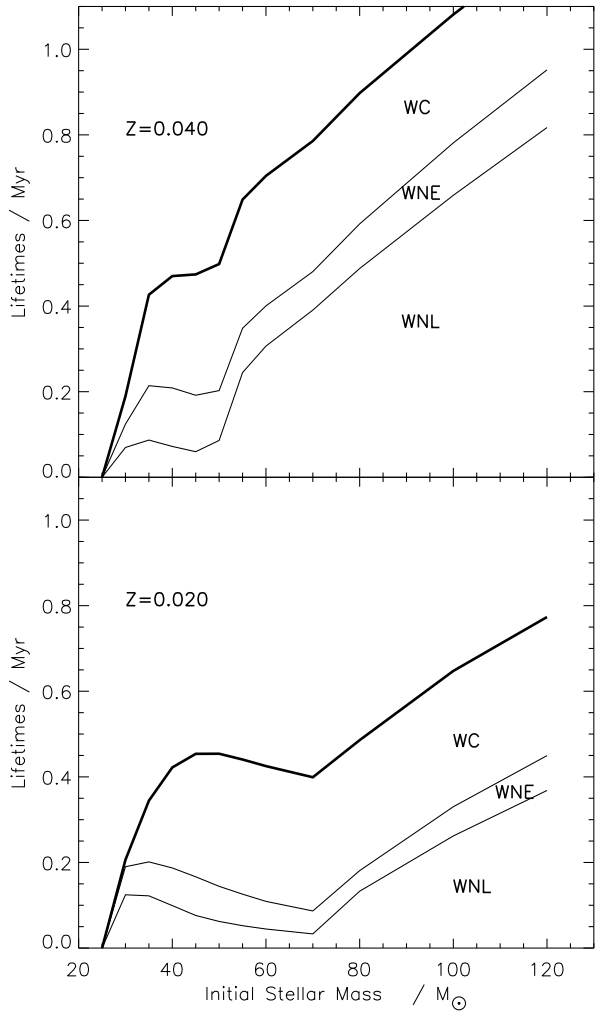}
\caption{The variation of the WR lifetimes with initial mass when the mass-loss rates are not scaled (scheme A). Each region is labelled but progresses from bottom to top by WNL, WNE and WC phase. The upper panel is for Z=0.04 and the lower panel is for Z=0.02.}
\label{wcwn1a}
\end{figure}
\begin{figure}
\includegraphics[angle=0, width=84mm]{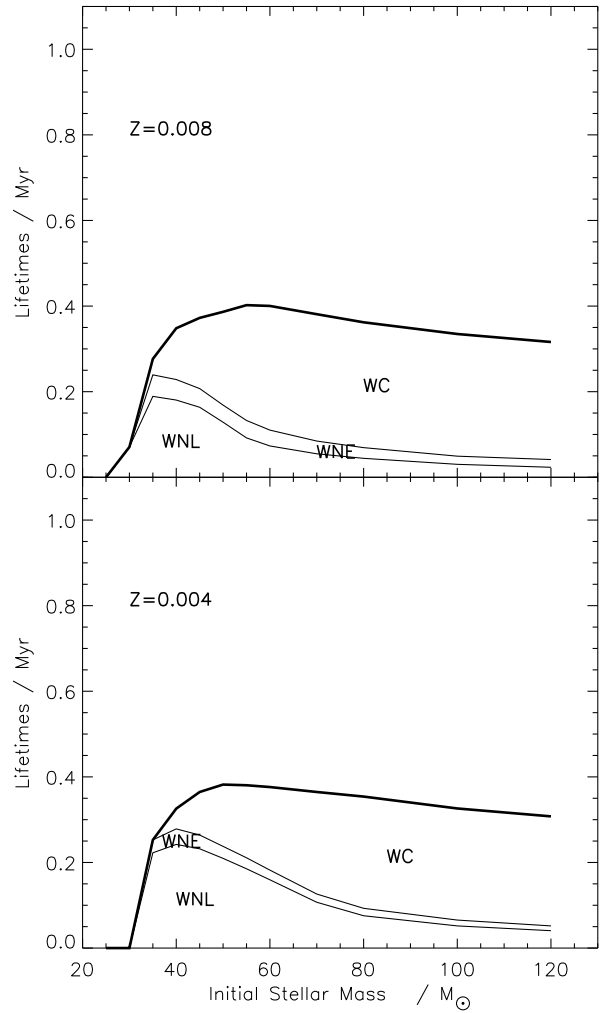}
\caption{Same as Fig.~\ref{wcwn1a}, but with metallicities Z=0.008 (upper panel) and Z=0.004 (lower panel).}
\label{wcwn1b}
\end{figure}

\begin{figure}
\includegraphics[angle=0, width=84mm]{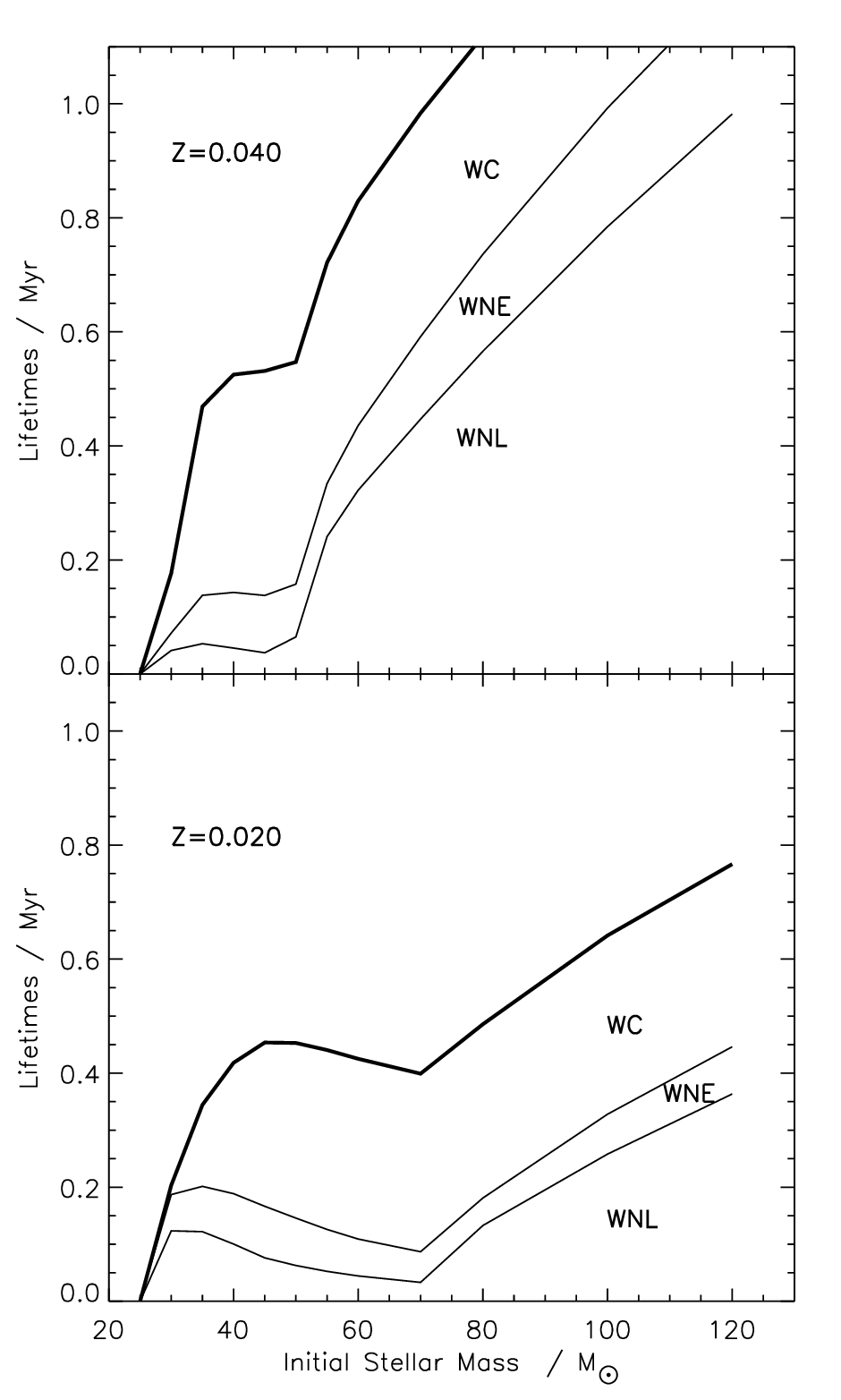}
\caption{Same as Fig.~\ref{wcwn1a} but with the mass-loss rates scaled according to scheme B.}
\label{wcwn2a}
\end{figure}

Figures~\ref{wcwn1a} and \ref{wcwn1b} show the times spent in the various phases when we do not scale \mdot\ with $Z$ (Scheme A). At higher metallicities, the WR lifetime is longer, because of larger pre-WR mass loss so that the WR phase is entered at an earlier stage resulting in a longer period of core helium burning during the WR phase. The kinks seen in the solar and twice solar $Z$ plots (Figs. \ref{wcwn1a} and \ref{wcwn2a}) are due the fact that stars with masses below the kink undergo a red supergiant phase, where mass is stripped and the surface H abundance drops below the limit set for WNL stars ($X_{\rm surface}$ $=$ $0.4$). However, the stars are still cool and some time passes before the stars are hotter than the temperature limit ($10^{4}\, {\rm K}$). For stars more massive than the kink, the stars spend most of their evolution at temperatures hotter than this limit and become WR stars as soon as the surface abundance requirement is reached.

\begin{figure}
\includegraphics[angle=0, width=84mm]{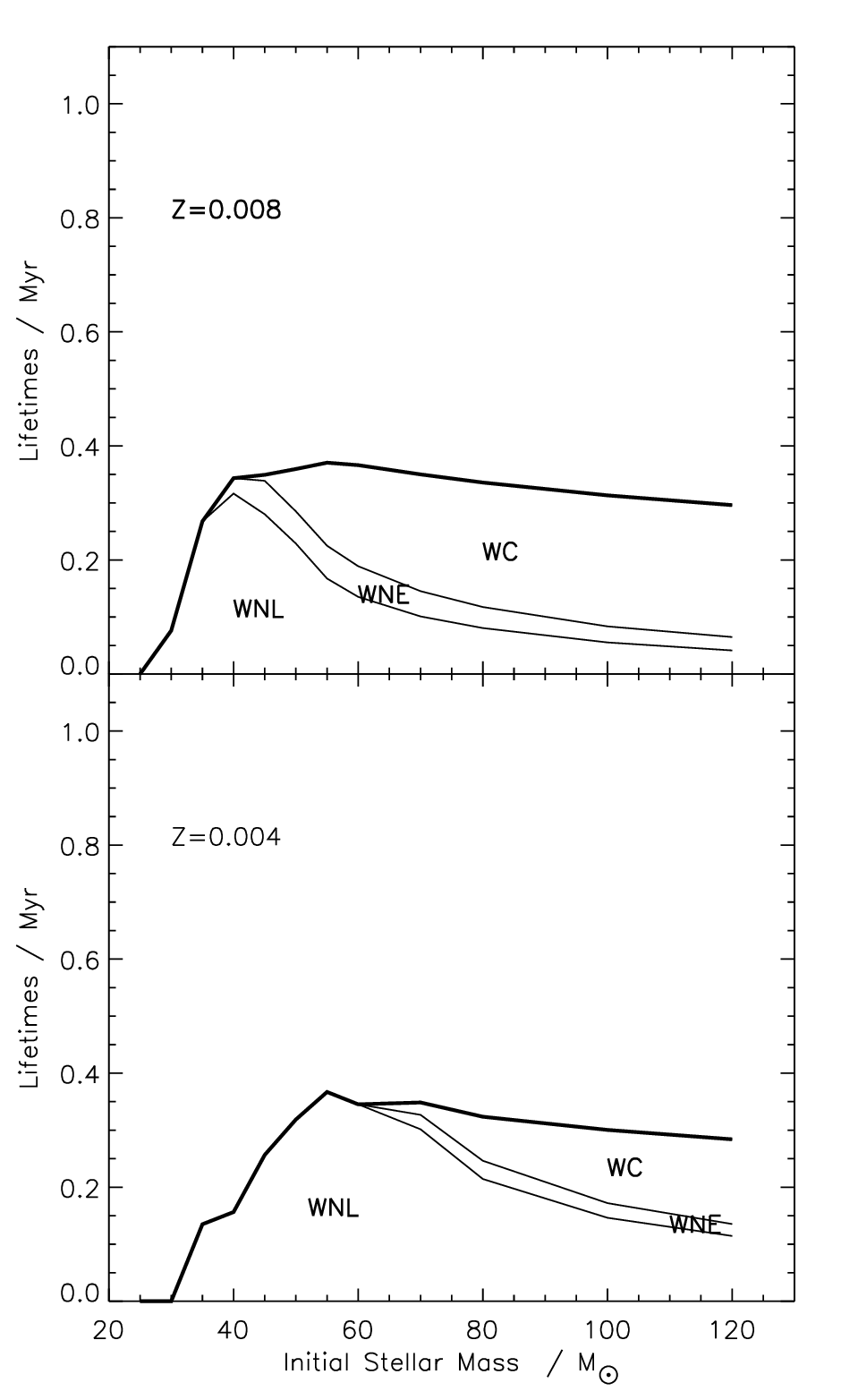}
\caption{Same as Fig.~\ref{wcwn1b} but with the mass-loss rates scaled according to scheme B.}
\label{wcwn2b}
\end{figure}

Figures~\ref{wcwn2a} and \ref{wcwn2b} show the times spent in the various WR phases when the predicted mass-loss scaling is included (Scheme B). The differences between the $Z$-scaled and unscaled models are relatively minor as far as the total WR lifetimes are concerned, although at $Z=0.04$ the lifetimes are somewhat longer in scheme B, as the WR stars are less massive and the nuclear reactions proceed at a slower rate, extending the lifetime. The reverse occurs at sub-solar $Z$. 

When we consider the lifetimes of the WR sub-phases, the mass-loss scaling affects how quickly nuclear burning products are exposed at the surface. At $Z=0.04$, we find that for stars below $50M_{\odot}$ the WN lifetime is reduced, while it is increased for stars above this value. The enhanced mass-loss leads to smaller He burning cores, so a longer time must pass before the He burning products are exposed at the stellar surface.

At $Z=0.008$ and $Z=0.004$ we find that the lower mass objects end their lives as WNL stars and do not progress any further along the WR evolutionary path. This is due to the weakening mass-loss stripping of a large fraction, but not all of the H. For more massive stars it leads to a reduction of the WC lifetime to the benefit of the WN lifetime. At even lower metallicities, the $\mdot-Z$ scaling leads to a complete absence of WC stars in our single star computations. 

\section{WC to WN populations ratio versus metallicity}
\label{ratio}

To evaluate whether the predicted $\mdot-Z$ scaling is similar to that in nature, we wish to compare our predictions to observations. We have compared the observed minimum initial mass required for the different WR phases \citep{masseyetal2000} as well as the observed ratio of type Ibc to type II SNe versus metallicity \citep{snevsZ} to our predicted values. We find that the inclusion of the predicted mass-loss scaling provides a better agreement with these observations, however the usefulness of these tests is limited by their large uncertainties. Fortunately, a more robust test is possible: the ratio of the WC versus WN star populations at different metallicities. We weight the WR lifetimes (Sect.~\ref{wrlifetimessection}) by an initial mass function \citep{KTG93} and predict WC/WN ratios that we compare to observed values. We implicitly assume a constant star-formation rate.

\begin{figure}
\includegraphics[angle=270, width=84mm]{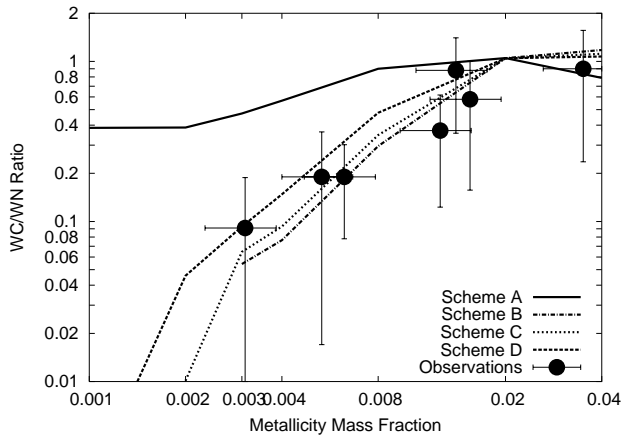}
\caption{The WC/WN ratio versus metallicity. Observed points are from \citet{wrdata1,wrdata2,wrdata3}. The error bars for the metallicity of the observed points reflect the uncertainty of converting from $\log_{10}({\rm O/H})+12$ and the uncertainty in the exact value of solar metallicity and therefore where the mass-loss rates should be scaled from.}
\label{allwcwn}
\end{figure}

\begin{figure}
\includegraphics[angle=270, width=84mm]{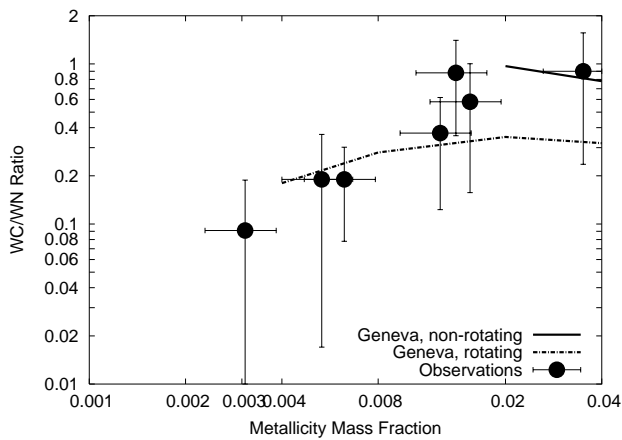}
\caption{Same as Fig.~\ref{allwcwn} but using the Geneva models.}
\label{allwcwn_genv}
\end{figure}

Figure~\ref{allwcwn} shows these predictions as well as the observed values. For the observed values the metallicity mass fraction was calculated from the $\log_{10}({\rm O/H})+12$ values by comparing them to the values from our sets of models. Because this process is ambiguous and as discussed above the position of solar metallicity for the mass-loss scaling is indistinct, we have assumed the metallicities are uncertain by 25 percent. In \citet{mm2005} a similar graph is presented where the x-axis was left as $\log_{10}({\rm O/H})+12$. In that plot an assumed solar value was taken rather than the composition of the stellar models. We here re-plot the Geneva model results in Fig.~\ref{allwcwn_genv} on the same scale as in Fig.~\ref{allwcwn}.

Despite the uncertainties in the observations it is clear that the observed WC/WN ratio decreases at lower $Z$. The models of \citet{mm2005} suggest that by mixing a population of rotating and non-rotating stars, agreement can be made between their models and the observations (Fig.~\ref{allwcwn_genv}). In Fig.~\ref{allwcwn} we show that agreement can be reached without any such assumption. All that is required is to scale the WR mass-loss rates with initial metallicity.


A complete evaluation of the relative importance of our WR mass-loss scaled models versus models including stellar rotation is difficult, as there are no 
contemporary non-rotating Geneva models available at lower metallicities. In addition, the Geneva models only include a WR metallicity scaling for supersolar 
models, despite the fact that the evidence used for this supersolar scaling comes from observations of WR stars in the low metallicity environment of the LMC. 
We nevertheless anticipate that including a WR $\mdot - Z$ scaling in rotating models will lead to an increase in the final WR masses (predicted by the Geneva code) 
and a decrease of the WC lifetimes, possibly even lowering the WC/WN ratio to values below those observed. 

In other words, Fig.~\ref{allwcwn} hints that as far as the WC/WN ratio is concerned, rotation is of secondary importance to employing the correct WR scaling of the mass-loss rates with metallicity. This assertion does not imply that rotation is not important for massive star evolution, as there is significant observational evidence for rotation-induced mixing (e.g. Howarth \& Smith 2001), but we note that this evidence is confined to the earlier stages of stellar evolution.
 
Finally, to check the sensitivity of our results to remaining uncertainties in the exponents of the scaling of the WR mass-loss rates, we study two extra scaling schemes listed in Table \ref{schemetable}. By comparing schemes B, C and D, we find that the rate at which the WC/WN ratio decreases is primarily determined by the exponent of the $\mdot-Z$ dependence during the WN phase. Although the error bars of the observed data-points are relatively large, agreement with the observations favours the $Z$-dependent 
schemes B and C. We note that it is only the quantity of the final WR mass that depends significantly on the on $\mdot-Z$ exponent during the WC phase.

\section{Summary, discussion \& conclusions}
\label{concl}

We have investigated how the predicted WR mass loss scaling with initial metallicity affects the output of evolutionary models. The inclusion of this scaling introduces important differences in WR masses and lifetimes, and produces a better agreement with observations, in particular the WC/WN ratio. Of course the physical effects of  rotation and binarity will be of additional importance.

The primary effects of rotation will be to decrease the minimum initial mass for WR stars and to boost the WR lifetimes, thereby increasing the total number of WR stars and the number of type Ibc SNe progenitors. Rotation has been investigated by \citet{mm2005}, which indicated that initially rapidly rotating stars have a lower WC/WN ratio than non-rotating stars. In other words, if rotation were included in our models the WC/WN ratios would decrease relative to the values presented here.

Binary stars will have similar effects to rotation in that they will lower the minimum initial mass for WR stars by providing extra opportunities for mass-loss via interactions. However, during the WR phase there will be little effect on the evolution, as WR stars have a smaller radius than during previous evolutionary phases, and the radius will commonly be smaller than the binary orbit. Therefore, we expect that more WR stars will exist in relation to other stellar types (such as red supergiants), but 
the WC/WN ratio will only be slightly affected.

So far we have not mentioned the importance of magnetic fields. \citet{magfields} model rotation and binary stars both with and without magnetic fields, and find significant differences in the behaviour of the stars. The most important effect of magnetic fields is to force a star to rotate as a solid body. This reduces the effect of rotation on the later stages of evolution as the star loses more angular momentum than when magnetic fields are not included and angular momentum is retained in the star's core. The study by \citet{magfields} comprises only a few stellar models concentrating on the question whether the WR stars can be the progenitors of long-duration GRBs. 
To obtain a more complete picture of the WC/WN ratio, the additional physical processes must be incorporated, however this would require a large number 
of stellar models, and is beyond the scope of this article. 

In closing, although both rotation and binarity have important implications for WR evolution, these are anticipated of only secondary relevance to ensuring the mass-loss metallicity dependence, as far as WR lifetimes, WC/WN ratios, and final masses of compact remnants are concerned. 

\begin{acknowledgements}
We thank the referee, Dr. W.-R. Hamann, for constructive comments that have helped improve the paper. 
JJE wishes to thanks the CNRS and the IAP. JSV wishes to thank RCUK for his Academic Fellowship.
\end{acknowledgements}

\end{document}